\documentclass{webofc}
\usepackage[varg]{txfonts}   %
\usepackage{inputenc}[utf8]
\usepackage{bm}
\usepackage{amssymb}
\usepackage{amsmath}
\usepackage{color}
\usepackage{subfigure}
\usepackage{cancel}
\usepackage{eucal}
\usepackage[normalem]{ulem}

\begin{document}
\title{Location of the QCD critical point predicted by holographic Bayesian analysis}

\author{\firstname{Mauricio} \lastname{Hippert}\inst{1}\fnsep\thanks{\email{hippert@illinois.edu}} \and
        \firstname{Joaquin} \lastname{Grefa}\inst{2,3} 
        \and 
        \firstname{T.~Andrew} \lastname{Manning}\inst{4}
        \and
        \firstname{Jorge} \lastname{Noronha}\inst{1} 
        \and
        \firstname{Jacquelyn} \lastname{Noronha-Hostler}\inst{1}
        \and
        \firstname{Israel} \lastname{Portillo~Vazquez}\inst{3}
        \and
        \firstname{Claudia} \lastname{Ratti}\inst{3}
        \and
        \firstname{Rômulo} \lastname{Rougemont}\inst{5}
        \and
        \firstname{Michael} \lastname{Trujillo}\inst{3}
}

\institute{Illinois Center for Advanced Studies of the Universe \& Department of Physics,
University of Illinois Urbana-Champaign, Urbana, IL 61801-3003, USA 
\and
  Department of Physics, Kent State University, Kent, Ohio 44243, USA
\and
Department of Physics, University of Houston, Houston, TX 77204, USA
\and 
National Center for Supercomputing Applications, University of Illinois Urbana-Champaign, Urbana, IL 61801, USA
\and 
Instituto de F\'isica, Universidade Federal de Goi\'as,
Av. Esperan\c{c}a - Campus Samambaia,
CEP 74690-900, Goi\^ania, Goi\'as, Brazil
}

\abstract{%
We present results for a Bayesian analysis of the location of the QCD critical point constrained by first-principles lattice QCD results at zero baryon density. We employ a holographic Einstein-Maxwell-dilaton model of the QCD equation of state, capable of reproducing the latest lattice QCD results at zero and finite baryon chemical potential. 
Our analysis is carried out for two different parametrizations of this model, resulting in confidence intervals for the critical point location that overlap at one sigma. 
While samples of the prior distribution 
may not even predict a critical point, or produce critical points spread around a large region of the phase diagram, posterior samples nearly always present a critical point at chemical potentials of $\mu_{Bc} \sim 550 - 630$ MeV. 
}
\maketitle
\section{Introduction}
\label{intro}

Exploring the QCD phase diagram is one of the major goals of experimental programs at RHIC and FAIR~\cite{Lovato:2022vgq}. 
While the hadronic and quark-gluon plasma (QGP) phases are smoothly connected at low values of the baryon density~\cite{Aoki:2006we}, where first-principles lattice QCD simulations are feasible, a first-order transition line is conjectured at high densities, starting at a second-order critical endpoint (CEP).  

Here, we aim to extrapolate knowledge from lattice QCD, available at lower baryon chemical potential $\mu_B$, to draw predictions for the QCD CEP expected at large values of $\mu_B$. 
For that, we employ a holographic model of the QCD equation of state, which is capable of reproducing lattice QCD results and is compatible with findings on QGP properties from the phenomenology of relativistic heavy-ion collisions ~\cite{Rougemont:2023gfz}. %
By using Bayesian inference tools, we perform a systematic scan over model realizations, selecting those that reproduce a set of lattice QCD constraints at zero density~\cite{Borsanyi:2013bia,Bellwied:2015lba} with a probability given by the respective error bars. 
We then compute predictions for the QCD CEP corresponding to each of the selected models to find an \textit{a posteriori} probability distribution for the CEP location.

\section{Holographic model}
\label{sec-model}

Our description of the QCD equation of state is based on the gauge/gravity correspondence, which allows us to use dual black holes in a 4+1 dimensional asymptotically anti-de Sitter bulk spacetime to describe the physics of a thermal, strongly coupled field theory sitting in the 3+1 Minkowski boundary of that geometry. %

More precisely, we employ a bottom-up Einstein-Maxwell-dilaton (EMD) model~\cite{Rougemont:2023gfz}, in which a Maxwell field $A^\mu$ is used to endow dual black holes with baryon number, while a dilaton scalar field $\phi$ is used to break conformal invariance and shape the renormalization group flow of the theory. The action of our theory is given by
 \begin{equation}
    S= \frac{1}{2\kappa_{5}^{2}}\int_{\mathcal{M}_5} d^{5}x\sqrt{-g}\left[R-\frac{(\partial_\mu \phi)^2}{2}-V(\phi)-\frac{f(\phi)F_{\mu\nu}^{2}}{4}\right],
\end{equation}
where $g$ is the determinant of the metric, $R$ is the Ricci scalar, $F_{\mu\nu} = \partial_\mu A_\nu - \partial_\nu A_\mu$ is the Maxwell field strength, and $V(\phi)$ and $f(\phi)$ are potentials which are tweaked to reproduce QCD physics. 
To assess the robustness of our results, we employ two different parametrizations of the potentials $V(\phi)$ and $f(\phi)$:
\begin{enumerate}
    \item \textit{Polynomial-hyperbolyc Ansatz (PHA):} A more traditional parametrization, similar to the one used in~\cite{Critelli:2017oub}:
\begin{equation}\label{eq:hyperpoly_V}
 V(\phi) = -12\cosh(\gamma\,\phi)+b_2\,\phi^{2}+b_4\,\phi^{4}+b_6\,\phi^{6},
\end{equation}
\begin{equation}
\label{eq:hyperpoly_f} 
 f(\phi) = \frac{\mathrm{sech}(c_{1}\phi+c_{2}\phi^{2}+c_{3}\phi^{3})}{1+d_{1}}+\frac{d_{1}}{1+d_{1}}\mathrm{sech}(d_{2}\phi).
\end{equation}

    \item \textit{Parametric Ansatz (PA):} A parametrization where parameters directly control plateaus and exponential slopes in the potentials~\cite{Hippert:2023bel}:
\begin{equation}\label{eq:parametric_V}
 V(\phi) = -12\cosh\left[\left(\frac{\gamma_1\,\Delta\phi_V^2 + \gamma_2 \,\phi^2}{\Delta \phi_V^2 + \phi^2}\right) \phi\right],
\end{equation}
\begin{equation}
\label{eq:parametric_f} 
 f(\phi) = 1 - (1-A_1) \left[\frac{1}{2} + \frac{1}{2}\tanh\left(\frac{\phi - \phi_1}{\delta \phi_1}\right)\right] +\\ 
 - A_1\left[\frac{1}{2} + \frac{1}{2}\tanh\left(\frac{\phi - \phi_2}{\delta \phi_2}\right)\right].
\end{equation}
\end{enumerate}

Holographic models of this kind have successfully reproduced lattice QCD results at intermediate temperatures $T \sim 100 - 500$ MeV at both vanishing and finite baryon density~\cite{Rougemont:2023gfz}. They can also predict a CEP in the QCD phase diagram and naturally describe the nearly inviscid nature of the QGP observed in high-energy heavy-ion collisions~\cite{Kovtun:2004de}.

\section{Bayesian analysis}
\label{sec-model}

We wish to scan over realizations of the model parametrizations above to generate an ensemble of models distributed according to the error bars on lattice QCD results. We thus employ Bayes' theorem to find the posterior probability over model parameters, $\vec\theta$, given the lattice QCD constraints $\vec d$:
\begin{equation}\label{eq:posterior}
    P(\vec\theta \,|\, \vec d) = \frac{P(\vec d \,|\, \vec\theta) \, P(\vec\theta)}{P(\vec d)} ,
\end{equation}
where $P(\vec\theta)$ is the prior probability distribution over model parameters, and we treat $P(\vec d)$, known as the evidence, as a normalization factor for the posterior. 

The likelihood $\mathcal{L}(\vec\theta) \equiv P(\vec d \,|\, \vec\theta)$ represents the probability of obtaining the lattice QCD results $\vec d$ given a set of model parameters $\vec\theta$. We model it as a Gaussian,
\begin{equation}
    \mathcal{L} 
    \sim \frac{1}{\sqrt{\det \Lambda}}\exp\left\{-\frac{1}{2}\sum_{i,j}\frac{p_i(\vec \theta)-d_i}{\sigma_i}\left[\Lambda^{-1}\right]_{ij}\frac{p_j(\vec\theta)-d_j}{\sigma_j}\right\}, 
\end{equation}
where $d_i$, $\sigma_i$ and $p_i(\vec\theta)$ represent, respectively, the $i$-th point in the lattice results under consideration, the corresponding error, and the prediction for that point given model parameters $\vec \theta$. The matrix $\Lambda_{ij}$ models correlations between different points, given by an extra parameter $\Gamma \in (-1,1)$, which measures correlations between neighboring points. 

To draw samples from the posterior  in Eq.~\eqref{eq:posterior}, 
we employ a Markov chain Monte Carlo (MCMC), in which parameter sets $\theta^{(n)}$ are randomly modified 
at each iteration to find $\theta^{(n+1)}$, in such a way that eventually $\theta^{(n\to\infty)}$ becomes distributed according to the target probability distribution $P(\vec \theta \,|\, \vec d)$.  In particular, we use differential evolution MCMC~\cite{2006S&C....16..239T}.
After a sufficiently large number of iterations, the equilibrium probability distribution $P(\vec \theta \,|\, \vec d)$ is achieved, and we can obtain samples of the posterior. 

As inputs for our Bayesian analysis, we take the latest lattice QCD results for the entropy density and baryon susceptibility at vanishing baryon density from the Wuppertal-Buddapest collaboration~\cite{Borsanyi:2013bia,Bellwied:2015lba}. 
More details on this analysis and the MCMC implementation can be found in the supplemental materials of Ref.~\cite{Hippert:2023bel}.

\section{Results} 

\begin{figure*}
\centering
\includegraphics[height=5cm,clip]{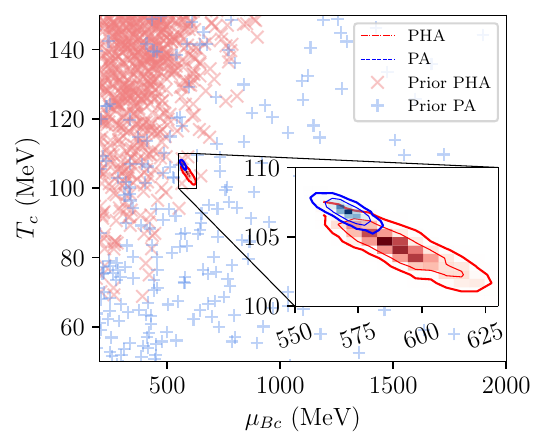}
\includegraphics[height=5cm,clip]{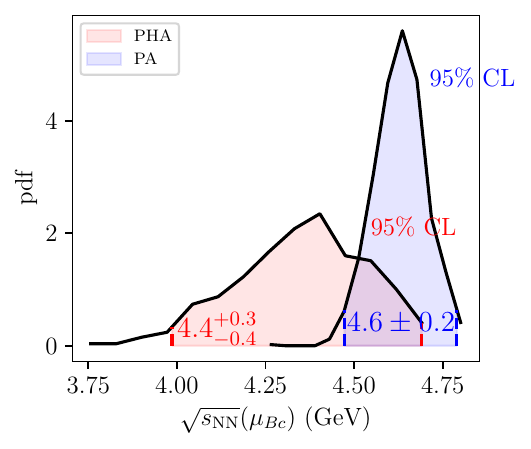}
\caption{Prior and posterior distribution for the CEP location in the PHA (red) and PA (blue) parametrizations. Left: Histograms for the critical temperature, $T_c$, and baryon chemical potential, $\mu_{Bc}$, and the corresponding $68\%$ and $95\%$ confidence levels in the posterior, together with critical point locations sampled from the prior (crosses). Right: Probability density function for the center-of-mass energy corresponding to $\mu_{Bc}$, according to the freeze-out line of Ref.~\cite{Vovchenko:2015idt}.  }
\label{fig:CEP}       %
\end{figure*}

Finally, we compute the predictions for the QCD CEP location in the samples of the posterior distribution we obtain and compute confidence levels for its location on the QCD phase diagram. 
For each parameter set or sample, we find the corresponding location of the CEP by following the procedure outlined in Ref.~\cite{Hippert:2023bel}. Results are shown in Fig.~\ref{fig:CEP}, where confidence levels for the critical temperature, $T_c$, and baryon chemical potential, $\mu_{Bc}$ are shown, alongside the posterior distribution for the corresponding beam energy, extracted from $\mu_{Bc}$ with the parametrization from Ref.~\cite{Vovchenko:2015idt}. CEP locations for the prior distribution are shown as crosses in the left panel. 

\section{Conclusions}
\label{sec-conc}

We have presented results for the first Bayesian analysis of the phase diagram of QCD constrained by first-principles lattice QCD results at zero baryon density. The posterior distribution of CEP locations was computed for two different parametrizations of a holographic EMD model. 
We find that imposing agreement with lattice QCD tightly constrains predictions for the QCD CEP location, which were spread all around the phase diagram in the unconstrained prior. Moreover, bands for the CEP location within each model overlap within one sigma, indicating the robustness of our results against parametrization choices. While 20\% of the prior predicts no CEP, a CEP is found in nearly all of the posterior, indicating that a CEP is statistically favored~\cite{Hippert:2023bel}.

\section*{Acknowledgements}

This material is based upon work supported by the National Science Foundation under grants No. PHY-2208724 and No. PHY-2116686 and in part by the U.S. Department of Energy, Office of Science, Office of Nuclear Physics, under Awards Number  DE-SC0023861 and DE-SC0022023. 
This work was supported in part by the National Science Foundation (NSF) within the framework of the MUSES collaboration, under grant number No. OAC-2103680. 
The authors also acknowledge support from the Illinois Campus Cluster, a computing resource that is operated by the Illinois Campus Cluster Program (ICCP) in conjunction with the National Center for Supercomputing Applications (NCSA), and which is supported by funds from the University of Illinois Urbana-Champaign. 
R.R. acknowledges financial support by the National Council for Scientific and Technological Development (CNPq) under grant number 407162/2023-2.

\bibliography{refs}

\end{document}